\documentclass{aastex}
\usepackage{emulateapj5}
\usepackage{epsfig}
\usepackage{rotate}
\newcommand{\ros}{{\sl ROSAT}}
\newcommand{\chan}{{\sl Chandra}}
\newcommand{\cxo}{{\sl CXO}}
\newcommand{\einst}{{\sl Einstein}}
\newcommand{\edot}{\dot{E}}
\newcommand{\fdeg}{\hbox{$.\!\!^{\circ}$}}
\begin{document}
\submitted{Submitted: April 12, 2000}
\title{ 
Variability of the Vela Pulsar-wind Nebula 
Observed with Chandra
}
\author{George G. Pavlov, Oleg Y. Kargaltsev, Divas Sanwal,
and Gordon P. Garmire}
\affil{The Pennsylvania State University, 
525 Davey Lab., University Park, PA 16802
\email{pavlov@astro.psu.edu}}
\begin{abstract}
The observations of the
pulsar-wind nebula (PWN) around the Vela pulsar with
the Advanced CCD Imaging Spectrometer 
aboard the
\chan\ X-ray Observatory, 
taken on 2000 April 30 and November 30, reveal its complex
morphology reminiscent of that of the Crab PWN.
Comparison of the two observations 
shows changes up to $30\%$ in the  surface
brightness of the PWN features. 
Some of the PWN elements show appreciable shifts, up to a few
arcseconds ($\sim 10^{16}$ cm),
and/or spectral changes.
To elucidate the nature of the observed variations,
further monitoring of the Vela PWN is needed.
\end{abstract}
\section{Introduction.}
X-ray and radio observations show that many of young pulsars
are enveloped by compact nebulae emitting a nonthermal spectrum
(e.g., Gaensler 2000).
It is commonly accepted that this emission is synchrotron radiation
from a relativistic pulsar wind shocked in the ambient medium.
The most famous and 
best studied example is the Crab pulsar-wind nebula (PWN).
High-resolution observations of the Crab PWN have revealed
its remarkably complex morphology in radio, optical and
X-rays (Bitenholz \& Kronberg 1992;
Hester et al.~1995; Weisskopf et al.~2000, and references
therein). Moreover, the 
multi-year studies
have shown complex variability of the Crab PWN. For instance,
the famous ``wisps'', discovered in optical observations,
show changes in their structure and brightness (Scargle 1969;
Hester et al.~1995), while the bright 
``knots'' change
their location and appearance between images separated by
only six days (Hester 1998). Recent 
observations with
the {\sl Chandra} X-ray Observatory
({\sl CXO}) have shown variations on a timescale of several
weeks (D.~Burrows \& J.~Hester, personal communication).
Greiveldinger \& Aschenbach (1999) analyzed 
five \ros\ 
observations of 1991--97 
and found monotonic changes 
in the surface 
brightness, some increasing and others decreasing,
at a rate of $\sim 2\%$
yr$^{-1}$.
Studying the PWN variability offers a unique opportunity
to understand the structure and dynamics of
the relativistic pulsar winds, elucidate the mechanisms
of PWN formation, evolution and interaction with the ambient
medium, and establish the properties of the relativistic
plasmas in PWNe. 
Thanks to the superb angular resolution of {\sl CXO},
a unique opportunity for such investigations is provided by the Vela PWN.

Observations of the 
Vela pulsar with \einst\ (0.1--4 keV band)
have shown that it
is embedded in a 
``kidney-bean'' 
nebula of  $\sim 2'$ size,
which emits a power-law spectrum with a photon index $\gamma =1.7\pm 0.2$
(Harnden et al.~1985).
The Vela PWN was further studied in soft X-rays
with {\sl EXOSAT} (\"Ogelman \&
Zimmermann 1989) and \ros\ 
(\"Ogelman, Finley, \&
Zimmermann 1993; Markwardt \& \"Ogelman 1998),
and it has also been detected at higher X-ray energies
with the
{\sl Birmingham Spacelab 2} (2.5--25 keV; Willmore et al.~1992)
and {\sl Compton Gamma-ray Observatory} (44--370 keV; 
de Jager, Harding, \& Strickman 1996).
\"Ogelman \& Koch-Miramond (1989)
claimed optical detection of 
a diffuse nebula around the pulsar, with a size of $40''$--$90''$
and V magnitude of 16--17,
which has not been confirmed by later observations
(Mignani et al.~2001).
No extended radio emission, at a level of 60 $\mu$Jy per $20''\times 10''$
beam at $\lambda=6$ cm, was found at the site of the X-ray PWN,
although a ridge ($3'\times 0\farcm 7$) of highly polarized radio emission
was detected at $\simeq 2'$ NE of the pulsar (Bitenholz, Frail, \&
Hankins 1991).

First \cxo\ observations of the Vela pulsar and its PWN 
(Pavlov et al.~2000; Helfand, Gotthelf, \& Halpern 2001; Pavlov et al.~2001;
Kargaltsev et al.~2001)
have shown a spectacular fine structure which resembles 
the Crab PWN --- 
arcs, jets,
knots,
 and diffuse cometary
tails (see Fig.~1).
The symmetry axis of the PWN image
(P.A.$=307^\circ\pm 2^\circ$), which can be interpreted as the
pulsar spin axis, is nearly parallel to the direction
of the pulsar proper motion (P.A.$=305^\circ$, Bailes et al.~1989).
This suggests that the
``natal kick'' of the pulsar was directed along
the rotation axis of the neutron star progenitor
(e.g., Lai, Chernoff, \& Cordes 2001).
The spectra of PWN elements
are power laws of different slopes, which vary from $\gamma\simeq 1.3$
for the NW jet and inner arc to $\gamma\simeq 1.7$
for the outer diffuse nebula (Kargaltsev et al.~2001).
The X-ray luminosity of the whole nebula in the 0.1--10 keV range
is $L_{\rm neb}=(6.0\pm 0.5)\times
10^{32}$
erg s$^{-1}$ (for $d=250$ pc --- Cha, Sembach, \& Danks 1999),
only $\sim 10^{-4}$
of the pulsar spin-down luminosity $\edot=6.7\times 10^{36}$ erg s$^{-1}$
(vs. 0.045 for the Crab PWN).

First indication on the Vela PWN variability was reported by
Helfand et al.~(2001) who analyzed two images obtained with
the \cxo\ High Resolution Camera 
on 2000 January 20 and February 21. They
 noticed a $5\%$ brightening of
the outer arc and suggested that it may be connected with
the large pulsar glitch of 2000 January 16 (Dodson, McCulloch, \&
Costa 2000).
In this Letter we report new results on X-ray variability
of the Vela PWN obtained with 
the \cxo\ Advanced CCD Imaging Spectrometer (ACIS;
Garmire et al.~2001).
\section{Observations}
Two ACIS observations 
of the Vela pulsar and
its PWN were carried out on 2000 April 30
and November 30
with exposure times 10,577 s and 18,851 s, respectively.
In  both observations the target was imaged
on the back-illuminated chip S3.
To image the whole PWN on one chip, the
pulsar was offset from the ACIS-S aimpoint by $-1\farcm 5$ 
along the chip row.
To reduce pile-up, we used
1/2 subarray,
with a frame
time of 1.5 s. 
For the analysis, we used the pipe-line processed Level 2 data
(versions 13.2 and 12.1 for the first
and second observations, respectively) and the CIAO software, v.2.0.
A detailed analysis of the PWN spectra and morphology
will be presented elsewhere (Kargaltsev et al.~2001).
Herein we 
concentrate on comparison of the two observations.

To compare the PWN structures,
it is
particularly important to co-align the images properly.
Since no X-ray point sources, other than the piled-up pulsar,
 are detected in the $8'\times 4'$
field of view, we have to rely upon the aspect reconstruction
which translates the actual event positions on the chip
into sky coordinates. For the processing software versions
used, the rms aspect offset (error of celestial location)\footnote{see
{\tt http://asc.harvard.edu/mta/ASPECT/cel\_loc/cel\_loc.html}}
is smaller than $0\farcs6$, although in some cases the offset can be
as large as $2''$. 
To estimate 
the errors in our case, we found centroids of radial distributions 
of pulsar counts, which correspond to
 $\alpha=8^{\rm h} 35^{\rm m} 20\hbox{$.\!\!^{\rm s}$}628$,
$\delta=-45^\circ 10' 34\farcs93$, and
$\alpha=8^{\rm h} 35^{\rm m} 20\hbox{$.\!\!^{\rm s}$}614$,
$\delta=-45^\circ 10' 35\farcs26$, for the first and second
image, respectively. The difference of these positions,
$0\farcs36$,
is comparable with our estimated centroiding error, $\simeq 0\farcs5$
(about one ACIS pixel).

The images of the central part
of the PWN
are shown in the
upper and lower panels
of Figure 1
for the first and second observation, respectively.
The white contours in the upper panel enclose 
PWN regions
chosen for the comparison with the second observation.
The regions numbered 1 through 7
enclose
well-defined bright features such as the {\it outer arc} (2a+2b)
with a brightened {\it spot} (1) at its apex,
{\it NW jet} (3), {\it inner arc} (4a+4b), {\it SE jet} (5),
{\it NE knot} (6), and {\it SW knot} (7).
In addition, to examine the
variability in the {\it diffuse emission}, we define regions 8a and 8b.
The white contours in the bottom panel are plotted
at the same positions as
those in the upper panel.
A visual comparison of surface brightnesses within the white
contours immediately shows substantial differences between
the two images --- e.g., the SW jet is brighter, the spot is
dimmer, and the NE knot is 
invisible in the second
image. Moreover, we see that some elements appear at different
locations --- the SW knot moved $2''$ westward, and the outer
arc shifted by, on average, $2''$ north-west from the previous positions.
Therefore, we defined new positions
of six PWN elements in the second image with blue contours
which 
 enclose the same areas as (and are congruent to)
 the corresponding white contours.
To quantify the changes in the surface brightness, we measured
the numbers of counts per unit area per unit time within the
white and blue contours for the first and second observations,
respectively (see Fig.~2), 
and found that the changes are indeed quite significant
for some elements ---
e.g., $32\%\pm 4\%$ for the SE jet, 
$-26\%\pm 2\%$ for the spot, and $-19\%\pm 2\%$ for the outer arc
(the uncertainties are $1\sigma$ statistical errors),
while they are comparable with statistical fluctuations for the others 
(e.g., for the {\it shifted} inner arc).
To visualize the spatial distribution of the brightness/morphology
change, we scaled the images to the same exposure time
(10,577 s), adaptively smoothed them with CIAO task {\tt csmooth}, 
 and subtracted the first
image from the second one. The difference image (Fig.~3) clearly
shows the displacements of the arcs
and the SW knot, brightening
of the SE jet, and dimming the spot and the NE knot.
It also shows some
brightening (dimming) of the diffuse emission at the NE
(SW) outskirts of the PWN and the large displacement, $\approx 8''$
toward SW,
and shape variation of the filamentary structure in the upper
right corner of the image ($75''$ NW from the pulsar). 

To evaluate 
systematic errors of the brightness changes, 
one should take into account that,
although the pulsar was imaged at almost the
same location on the chip in the two observations,
the PWN regions are imaged on different
sites because
the roll angles were different
($260\fdeg 3$ and $55\fdeg 2$). Since the CCD 
quantum efficiency 
varies over the chip\footnote{see
{\tt http://asc.harvard.edu/ciao/wrkshp/esa1.pdf}}, this may lead to artificial changes of
brightness. To examine 
this effect, we employed two sets of on-orbit
calibration exposures (28 ks and 35 ks,
at dates close to those of our 
observations), during which
the ACIS chips were illuminated by on-board radioactive
X-ray sources.   For each of the calibration datasets,
we measured
the surface brightnesses of eight domains 
on the chip where four PWN elements (regions 1, 2b, 4a and 5)
were imaged
in the first and second observations of the Vela PWN.
The differences of the brightnesses of the domains corresponding
to the same PWN elements, both in the separate calibration
datasets and between the two datasets,
 are about 2\%--3\%, comparable to the statistical
errors. This puts an upper limit of $3\%$ on both spatial and temporal
variability of quantum efficiency within the chip area
($\sim 1'\times 1'$) where the
 bright PWN elements were imaged.
As an independent test, which additionally accounts for the effect of dither 
on the 
exposure times,
we constructed exposure maps (e.g.,
Davis 2001) for the two
PWN observations, for an energy of 1 keV close to the maxima
of the count rate spectra. Inspection of these maps shows
that the effective area varies by about 20\% over
the whole $8'\times 4'$ field-of-view, but these large variations
are associated with the boundaries between the four chip nodes.
Since the PWN elements under investigation are all imaged within the
same node 1,
 the variations
are much smaller, $\approx 2\%$.
Thus, 
the large, $\sim 20\%$--30\%, brightness
changes we detected is not an instrumental effect --- they characterize 
real PWN variations.

We have also measured the mean surface brightnesses
in the two
observations within a 
$30''$ radius
circle (centered on the pulsar).
To take into account the above-mentioned nonuniformity,
we divided the original images over the exposure maps
and found a 
change of $-3.0\%\pm 0.5\%$.
Although this difference looks statistically significant, its 
magnitude is comparable with possible systematic errors
(e.g., caused by inaccuracy of the monoenergetic exposure maps we used).

Comparison of the two observations allows one, in principle, to check
whether the pulsar luminosity has changed in 7 months. 
A change of the luminosity could be caused
by 
the strong glitch of 2000 January 16 because glitch
effects can manifest themselves on time scales
from weeks to years, depending 
on the depth where the energy release
occurred. We measured the radial count distribution of the piled-up
pulsar image and found that its very central part,
within a $1''$ radius,
became brighter by $30\%\pm 5\%$,
whereas the brightness did not change
at larger radii.
In obvious contradiction with this result, 
the difference in the pulsar countrates,
$-8\%\pm 9\%$,
estimated from the 
``trailed images'' (one-dimensional images
formed during the frame read-outs) 
is statistically insignificant.
 Since the trailed images do not suffer from
the pile-up, we consider the latter result more reliable.
The apparent variation of the central part of the 
pulsar image is likely caused by a small fluctuation
of observing conditions (e.g., focal plane temperature),
which may lead to a considerable change in the nonlinear regime 
associated with strong pile-up.  It should also be mentioned
that the piled-up images are very 
asymmetric. In particular,
in both observations we see a ``blob'' of  $0\farcs7$ radius at a distance
of $1\farcs5$ from the center of the pulsar image (see Fig.~3),
at an angle of about $22^\circ$ from the $Z_+$ axis of the spacecraft.
The blob is most likely due to a tilt between the telescope
mirrors in the innermost shell (Jerius et al.~2001). 
It can hardly be seen when there is no pile-up
because it is much fainter than the core of the point
source image, but it appears quite bright in a strongly
piled-up image because the pile-up reduces the core brightness. 

In addition to the shifts and brightness changes, we have examined
the spectral changes of the PWN elements.
To characterize the spectra, we use the hardness ratio $h_{1.3}$,
defined as the ratio of counts above and below 
$E=1.3$ keV\footnote{
Being defined as a linear function of pulse-height amplitude 
(see {\tt http://asc.harvard.edu/cal/Links/Acis/acis/Cal\_prods/gaincti/06\_19\_00/cti.html}), $E$ corresponds to 
the most probable photon energy for a recorded PHA.}.
For the comparative analysis, this empirical quantity is more convenient
than spectral fitting parameters because it does not depend on the
spectral model, and does not suffer from uncertainties
of the detector spectral response. 
We see some evidence of hardness changes,
up to $+29\%\pm 8\%$ in the SE jet (see Fig.~2),
but their statistical significance,
$<3.6\sigma$,
is not as high as that of the brightness changes.
\section{Discussion.}
The comparison of the two ACIS observations taken 7 months apart shows
that the Vela PWN is by no means static --- we have detected 
considerable brightness changes and/or shifts of some PWN elements,
which are likely accompanied by spectral changes.
With only two observations carried out, we cannot trace these variations
to establish their time scales and 
velocities.
In particular,
it is hard to conclude whether the apparent
shifts of the arcs and the SW knot are associated with 
steady motion of matter during the 7 months or are manifestations
of wave phenomena, or some instabilities,
 in the post-shock relativistic
plasma. The fact that the NE knot has disappeared and the other PWN 
elements have changed their brightness implies that the changes
are more complicated than simple motion of
plasma inhomegeneities.

The remarkable similarity
of the Vela and Crab PWN morphologies suggests 
that their variabilities are driven by similar physical processes.
The main distinction
between the two PWNe is in their sizes and energetics. In particular,
the physical size of the Vela PWN, $\sim 0.1$ pc at $d=250$ pc,
is an order of magnitude smaller than that of the Crab PWN, in
rough correspondence with the scaling,
$r_s\propto \dot{E}^{1/2}$, 
for the shock radius in an ambient medium with a given pressure.
This correspondence indicates that typical pressures,
$p\sim \dot{E}/4\pi c r_s^2\sim 10^{-9}$ erg~cm$^{-3}$, are of the same 
order of magnitude, which is in agreement with the
 close values of their equipartition magnetic
fields, $B\sim 10^{-4}$ G (Kargaltsev et al.~2001). The similarity
of physical parameters of the relativistic plasmas in the two PWNe
suggests that typical velocities of MHD waves,
presumably responsible for 
some of the observed variabilities,
are also similar.
In a relativistic magnetized plasma 
these velocities 
depend on the mean
energy density 
$\varepsilon$, magnetization parameter
$\sigma = B^2/(4\pi\varepsilon)$,
and the angle
$\theta$ between the wavevector and the magnetic field
(e.g., Gedalin 1993). For the conditions expected in the Crab and Vela PWNe,
they vary 
from very low values,
$\approx (c/2)\sqrt{3\sigma}|\cos\theta|$,
 for the Alfven and slow magnetosonic waves 
at $\sigma\ll 1$,
to $\simeq (0.6$--$0.7) c$ for the 
fast magnetosonic wave at $\sigma\simeq 1$. 
The fastest speeds observed for the Crab PWN wisps, about $0.5 c$
(Hester 1998), 
are thus close to the upper end of the expected velocity range.
If
plasma perturbations can propagate with similar
velocities in the Vela PWN, the observable shifts $\sim 1''$--$2''$
can occur in just few days, so that the 7 months between our observations
may not be a representative time period to investigate the variations.

The different behavior of the Vela PWN elements suggests
 that various processes are responsible
for the PWN variability --- e.g., the apparent motion of the arcs
can be caused by wave processes, 
while the brightness variations of the jets can be associated
with large-scale inhomogeneities of the polar outflows.
A variety of mechanisms have been invoked to explain the
Crab PWN variability --- e.g.,  formation of MHD shock waves
in slightly inhomogeneous wind streams (Lou 1998), instabilities
driven by synchrotron cooling in the flow (Hester 1995), 
nonlinear Kelvin-Helmholtz instabilities in the equatorial
plane of the shocked wind (Begelman 1999), or cyclotron instabilities
of ion rings (Spitkovsky \& Arons 2000). To investigate the actual
roles of such processes in the Vela PWN, its behavior should be monitored
with \cxo\ --- other X-ray missions do not 
have sufficient angular resolution,
while observations outside the X-ray band are extremely difficult, if
possible at all, because of the PWN faintness in the optical and radio.
Moreover, the Vela PWN is more suitable for studying
brightness and, particularly, spectral variabilities with the \cxo\
ACIS because it is not as bright as Crab, and its ACIS images
do not suffer from pile-up which strongly complicates the
data analysis. Thus, future \cxo\ observations of the Vela PWN
will provide the unique opportunity to elucidate the physical
properties of the relativistic plasmas in PWNe.
\acknowledgements
We are grateful to Leisa Townsley, George Chartas and Slava Zavlin
for useful discussions. This work was supported by SAO grant GO1-2071X
and NASA grant NAS8-38252.

\begin{figure*}
\epsscale{1.2}
\vspace*{-6cm}
\hspace*{-3cm}
\plotone{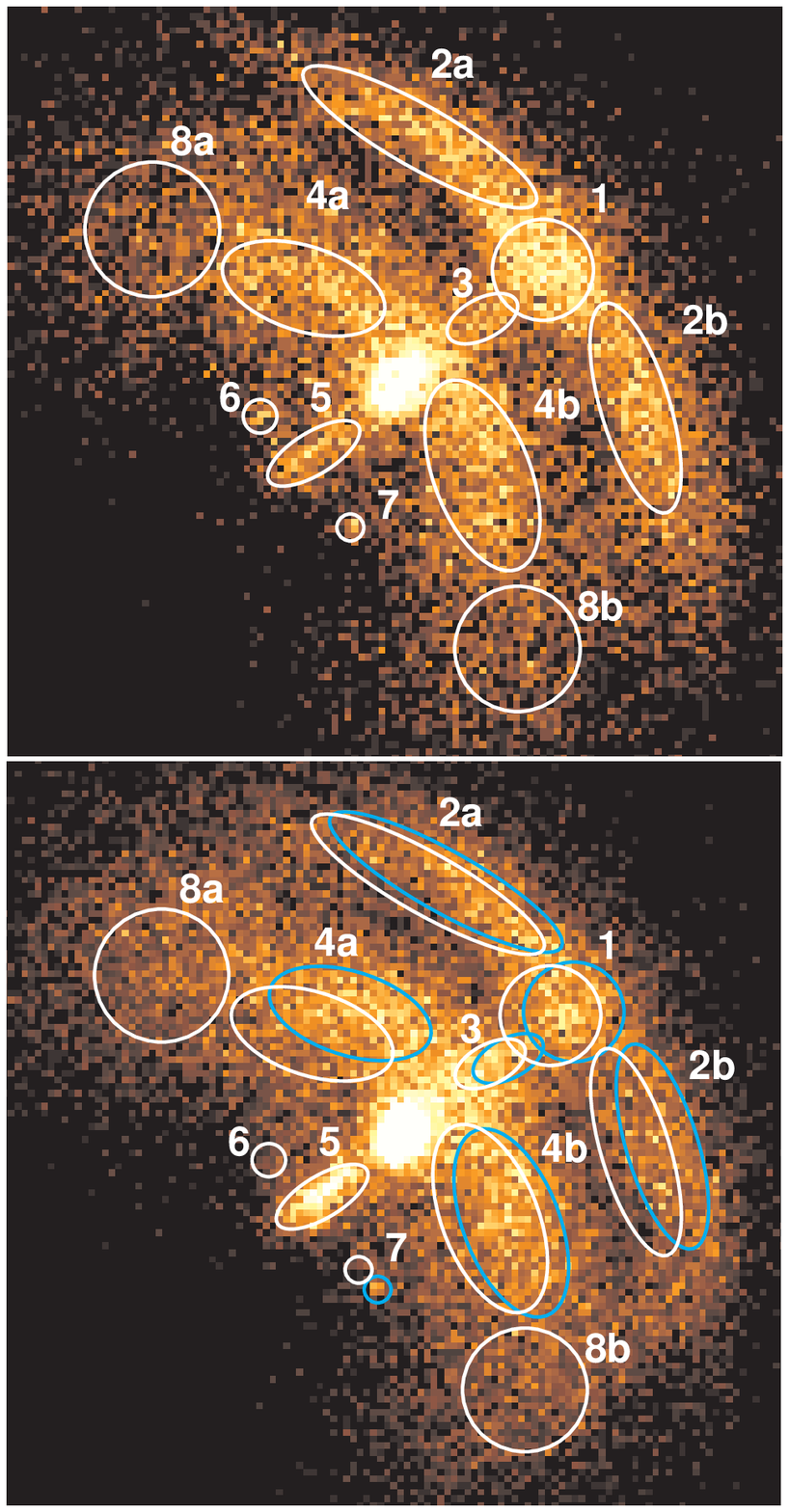}
\vskip -4cm
\caption{ACIS-S images
of a central part
($57''\times 55''$) of the Vela PWN of 2000 Apr 30 (top) and Nov 30 (bottom).
The pixel size is $0\farcs492$.
The white contours in the top panel define the PWN elements in the
first image. In the bottom panel, the white contours correspond
to the contours in the top panel, while the blue contours demonstrate
the displaced nebular elements in the second
observation.
The brightest spot is the Vela pulsar.
}
\end{figure*}
\begin{figure*}
\vspace*{-2cm}
\hspace*{-2cm}
\plotone{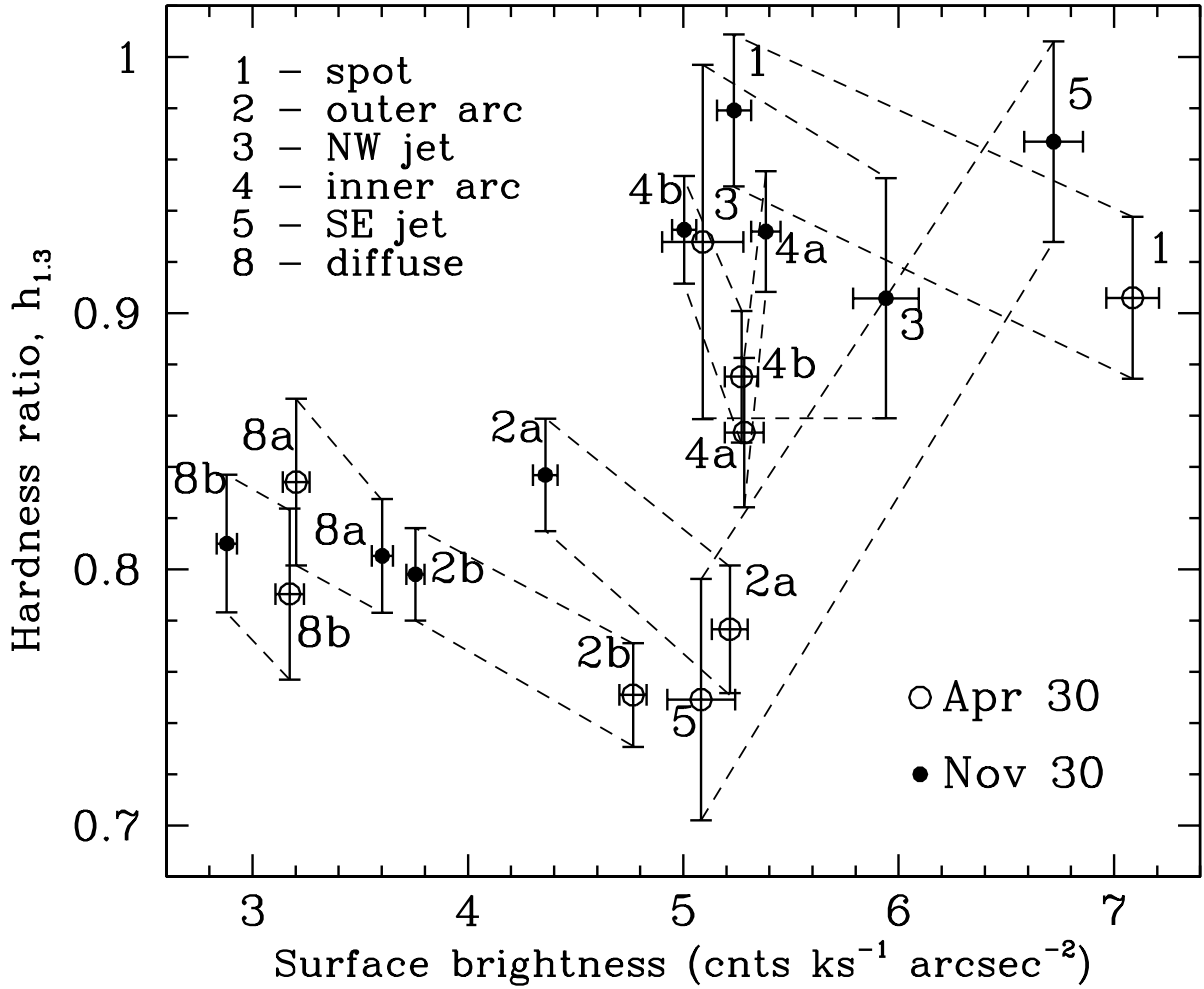}
\vskip -10cm
\caption{
Surface brightness vs.~hardness ratio 
$h_{1.3}$ 
for different PWN regions for the two
ACIS-S observations.
 For the second observation,
the values for the regions 1--4
 were calculated within the blue contours in the bottom panel of Fig.~1.
(The differences
would be even larger if exactly the same regions were used for the two
observations.)
}
\end{figure*}
\begin{figure*}
\epsscale{0.8}
\vspace*{-5cm}
\plotone{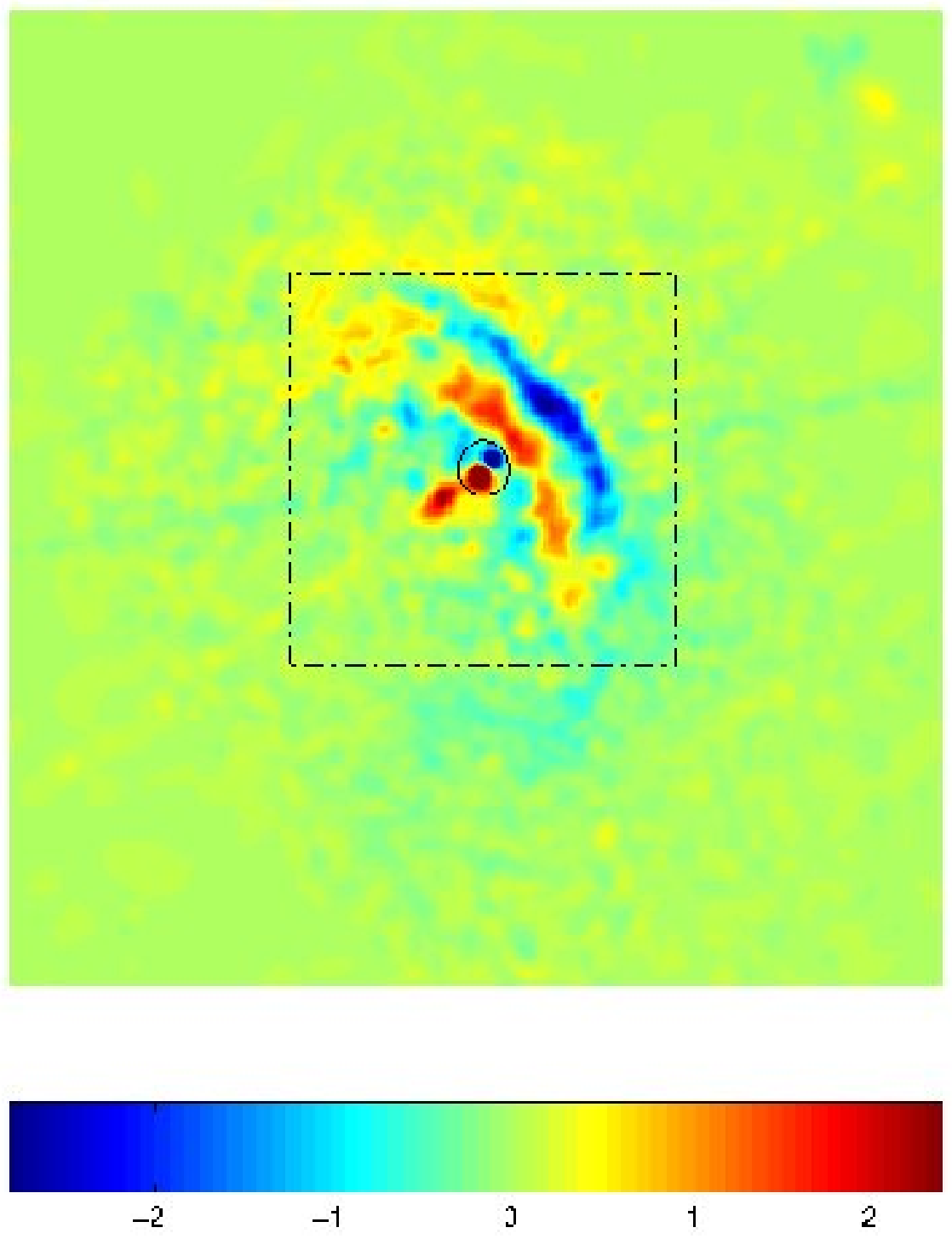}
\vskip 1cm
\caption{
The difference image ($142''\times 142''$) of the Vela PWN.
The color scale is in counts ks$^{-1}$ arcsec$^{-2}$.
The dashed rectangle corresponds to the size of images in Fig.~1.
The blue and red structures within the $3\farcs5$ circle at the
center correspond to the ``blob'' in the pulsar image in the first
and second observation, respectively (see text for details).
The blue and yellow linear structures are the pulsar trailed images.
}
\end{figure*}

\begin{thebibliography}{}
\bibitem{}
Bailes, M., 
Reynolds, J.~E., Manchester, R.~N., Kesteven, M.~J.,
\& Norris, R.~P.
1989, ApJ, 343, L53

\bibitem[Begelman 1999]{beg99} 
Begelman, M.~C. 1999, \apj, 512, 755

\bibitem{}
Bitenholz, M.~F., \& Kronberg, P.~P. 1992, ApJ, 393, 206

\bibitem[Bitenholtz et al. 1991 ]{}
Bitenholz, M.~F., Frail, D.~A., \& Hankins, T.~H. 1991, ApJ, 376, L41

\bibitem{}
Cha, A., Sembach, K.~M. \& Danks, A.~C. 1999, ApJ, 515, L25

\bibitem{}
Davis, J.~E. 2001, ApJ, 548, 1010

\bibitem[de Jager et al. 1996]{dej96} 
de Jager O.~C., Harding, A.~K., Strickman, M.~S. 1996, \apj, 460, 729

\bibitem{}
Dodson, R.~G., McCulloch, P.~M., \& Costa, M.~E. 2000, IAU Circ. 7347 

\bibitem{}
Gaensler, B.~M. 2001, in Young Supernova Remnants, eds. S.S. Holt \& U. Hwang,
AIP, New York (in press; also astro-ph/0012362)

\bibitem{}
Garmire, G.~P., et al. 2001, ApJS, in preparation

\bibitem{}
Gedalin, M. 1993, Phys.~Rev. E, 47, 4354

\bibitem[Greiveldinger \& Aschenbach (1999)]{gre99} 
Greiveldinger, C., \& Aschenbach, B. 1999, \apj, 510, 305

\bibitem[Harden et al. (1985)]{har85} 
Harden, F.~R., Grant, P.~D., Seward, F.~D., \& Kahn, S.~M.
 1985, \apj, 299, 828

\bibitem[Helfand et al,  2000]{hel00} 
Helfand, D.~J., Gotthelf, E.~V., \& Halpern, J.~P. 2001,
ApJ, accepted; astro-ph/0007310

\bibitem[Hester et al. (1995)]{hes95} 
Hester, J.~J., et al. 1995, \apj, 448, 240

\bibitem{}
Hester, J.~J. 1998, in Neutron Stars and Pulsars: Thirty Years
after Discovery, eds. N.~Shibazaki, N.~Kawai, S.~Shibata and T.~Kifune,
Univ.~Acad.~Press (Tokio), p.431

\bibitem{}
Jerius, D., et al. 2001, SPIE, in press ({\tt http://asc.harvard.edu/cal/Hrma/hrma/psf/index.html})

\bibitem[Kargaltsev et al, 2000]{kar00} 
Kargaltsev, O.~Y., Pavlov, G.~G., Sanwal, D., Garmire, G.~P., \& Zavlin,
V.~E. 2001, ApJ, in preparation

\bibitem{}
Lai, D., Chernoff, D.~F., \& Cordes, J.~M. 2001, ApJ, 549, 1111

\bibitem[Lou 1998]{lou98}
Lou, Y. 1998, \mnras, 294, 443

\bibitem[Markwardt \& \"Ogelman, 1998]{mar98} 
Markwardt, C.~B., \&  \"Ogelman, H. 1998, Mem. Soc. Astr. Ital., 69, 927

\bibitem{}
Mignani, R., et al. 2001, in preparation

\bibitem[\"Ogelman, Finley, {\&} Zimmerman (1993)]{oge93} 
\"Ogelman, H., Finley, J.~P., \& Zimmerman, H.-U. 1993, Nature, 361, 136

\bibitem[\"Ogelman \& Koch-Miramond 1989 ]{oge89}
\"Ogelman H. \& Koch-Miramond, L. 1989, \apj, 342, L83

\bibitem{}
\"Ogelman, H., \& Zimmerman, H.-U. 1989, A\&A, 214, 179

\bibitem{}
Pavlov, G.~G., Sanwal, D., Garmire, G.~P., Zavlin, V.~E., Burwitz, V.,
\& Dodson, R.~G. 2000, AAS Meeting 196, \#37.04

\bibitem{}
Pavlov, G.~G., Zavlin, V.~E., Sanwal, D., Burwitz, V., \& Garmire, G.~P.
2001, ApJ Lett, in press (astro-ph/0103171)

\bibitem[Rees \& Gunn 1974]{ree74}  
Rees, M. J., Gunn, J. E. 1974, \mnras, 167, 1

\bibitem{} 
Scargle, J.~D. 1969, ApJ, 156, 401

\bibitem[Spitovsky \& Arons (2000)]{spi00} 
Spitkovsky, A., \& Arons, J. 2000, in Pulsar Astronomy - 2000 and Beyond,
eds. M.~Kramer, N.~Wex, and R.~Wielebinski,
ASP Conference Series, v.202, p.507

\bibitem{}
Weisskopf, M.~C., et al. 2000, ApJ, 536, L81

\bibitem[Willmore et al. 1992]{wil92} 
Willmore, A.~P., et al. 1992, \mnras, 254, 139

\end{thebibliography}
\end{document}